\documentclass[showpacs,preprintnumbers,amsmath,amssymb,prb]{revtex4}

\usepackage{graphicx}
\usepackage{dcolumn}
\usepackage{bm}

\begin{document}

\title{Critical temperature and low-energy excitations in gapped spin systems with defects}

\author{F.\ D.\ Timkovskii$^1$}
\email{philippinho@yandex.ru}
\author{A.\ V.\ Syromyatnikov$^{1,2}$}
\email{asyromyatnikov@yandex.ru}
\affiliation{$^1$National Research Center "Kurchatov Institute" B.P.\ Konstantinov Petersburg Nuclear Physics Institute, Gatchina 188300, Russia}
\affiliation{$^2$St.\ Petersburg State University, 7/9 Universitetskaya nab., St.\ Petersburg, 199034
Russia}

\date{\today}

\begin{abstract}

We discuss theoretically the magnetically ordered phase induced by magnetic and nonmagnetic impurities in three-dimensional and quasi-low-dimensional systems with singlet ground states separated by a gap from excited triplet states. Using ideas of the percolation theory, we estimate the transition temperature $T_N(n)$ to the N\'eel phase at a small concentration $n$ of defects, derive the density of states of low-energy elementary excitations, and examine the contribution of these excitations to the specific heat and magnetization. Our expressions for $T_N(n)$ and for the specific heat describe well available experimental findings obtained in various appropriate systems: spin-$\frac12$ dimer materials, spin-ladder compounds, spin-Peierls and Haldane chain materials. However, our expression for $T_N(n)$ differs considerably from many of those proposed before. 

\end{abstract}


\maketitle

\section{Introduction}

Spin systems with singlet ground states separated by a gap from lowest (triplet) excitations have attracted much attention recently both experimentally and theoretically. Particular examples of such objects include systems containing spin-$\frac12$ dimers which are weakly coupled by three-dimensional interactions, spin-$\frac12$ ladders, spin-Peierls dimerized chains, and integer-spin Haldane chains.

It is well established that a magnetic or a non-magnetic impurity induces in these systems a local magnetic moment and a magnetically ordered cloud arises around the defect. \cite{ladder1,ladder2,ladder3,bobroff,haldane1,haldane2,haldane3,Oosawa2002,Oosawa2003,peierls1,peierls2} The staggered magnetization in the cloud drops off exponentially with the distance beyond the volume whose shape and size is determined by the ground-state properties of the pure host system. An RKKY-like effective interaction arises between the induced magnetic moments via these clouds, or, equivalently, via the gapped bulk excitations exchange (see also below). In host systems on bipartite lattices with commensurate spin correlations, this effective coupling is non-frustrated and it leads to a N\'eel magnetic order at small enough temperature $T_N(n)$ at finite impurities concentration $n$ (the phenomenon of the "order-by-disorder" type). Then, the disorder-induced magnetically ordered part of the system produces gapless excitations inside the singlet-triplet gap. To the best of our knowledge, these excitations have not been discussed analytically so far. 

One of the aims of the present paper is to fill up this gap. We demonstrate below that ideas of the percolation theory are very useful in solving this problem. We demonstrate in Sec.~\ref{excitation} that the disorder-induced band of excitations consists of two parts: the low-energy part is governed by long-wavelength propagating antiferromagnetic spin waves above which localized states appear. 
We show in Sec.~\ref{magnspecheat} that these excitations determine the behavior of the staggered magnetization and the specific heat at $T\ll T_N(n)$. 

Besides, we scrutinize below previous estimations of the N\'eel temperature $T_N(n)$. It was shown in Refs.~\cite{bobroff,Fabrizio1999,Melin2000} that various spin-ladder, spin-Peierls dimerized chain, and spin-1 Haldane chain materials show a linear dependence of $T_N(n)$ in a range of small $n$. An exponential dependence $T_N(n)\propto e^{-c/n}$ is frequently used \cite{peierls1,f3} in quasi-1D materials, which, however, requires unrealistic parameters $c$ to fit experimental data. \cite{bobroff} In contrast, we demonstrate in Sec.~\ref{neeltemp} that the dependence of the N\'eel temperature on $n$ is more complicated which can give a linear-like behavior in a certain range of $n$. We find that $T_N(n)\propto e^{-c/n^{1/3}}$ as it was estimated in Ref.~\cite{f2}, where, however, the constant $c$ was not obtained. It is shown in Sec.~\ref{experiment} that our formulas for $T_N(n)$ and the specific heat describe well existing experimental data in a variety of relevant compounds.

Sec.~\ref{conc} contains a summary and our conclusion.

\section{N\'eel temperature}
\label{neeltemp}

We adopt in our theoretical discussion ideas proposed in Refs.~\cite{Korenblit1973,Korenblit,shenderrev} for disordered ferromagnets. For definiteness, we consider below the spin-$\frac12$ dimer system on a cubic lattice whose Hamiltonian has the form
\begin{equation}
\label{ham}
{\cal H} = \sum_i	
\left( {\cal J}{\bf S}_{({\bf r}_i,1)}{\bf S}_{({\bf r}_i,2)} 
+ 
\sum_{j=x,y,z}	
J_j \left({\bf S}_{({\bf r}_i,1)}{\bf S}_{({\bf r}_i+{\bf e}_j,1)} 
+ {\bf S}_{({\bf r}_i,2)}{\bf S}_{({\bf r}_i+{\bf e}_j,2)}\right)
\right),
\end{equation}
where ${\bf S}_{({\bf r}_i,q)}$ is the spin $q$ ($q=1,2$) from the dimer at the lattice site ${\bf r}_i$, ${\cal J}>0$ is the intradimer exchange coupling constant, and $J_{x,y,z}>0$ are exchange coupling constants between spins from neighboring dimers along the corresponding directions. A generalization is straightforward of the results obtained below to other relevant spin models.

The effective exchange coupling between two induced spins inside the host system with the gapped spectrum of elementary excitations $\varepsilon_{\bf k} = \Delta\sqrt{1 + \xi_x^2k_x^2 + \xi_y^2k_y^2 + \xi_z^2k_z^2}$ is related to the static spin correlation function and has the form (see, e.g., Refs.~\cite{ladder1,MikeskaGhosh2004})
\begin{equation}
\label{jeff0}
J({\bf r}) 
= 
C \int \frac{H_m^2}{\varepsilon_{\bf k}} e^{i{\bf kr}} d{\bf k}
=
\frac{(4\pi)^2 C}{3} \frac{H_m^2}{V_\xi\Delta}\frac{1}{R} K_1( R )
\end{equation}
 at $r\gg1$, where $\Delta=\sqrt{{\cal J}({\cal J}-2 (J_x+J_y+J_z))}$ is the gap value, $\xi_{x,y,z}^2 = {\cal J}J_{x,y,z}/\Delta^2$, $C$ is a constant of the order of unity, $H_m=J_x+J_y+J_z$ is the local molecular field made by an impurity in the host system, $V_\xi = \frac{4\pi}{3} \xi_x\xi_y\xi_z$, ${\bf R} = \left( \frac{x}{\xi_x}, \frac{y}{\xi_y}, \frac{z}{\xi_z} \right)$, $K_1(R)$ is the modified Bessel function, and we omit the sign depending on whether or not the couple of spins belong to the same sublattice. At $R\ll1$, $K_1(R)\propto1/R$. Eq.~\eqref{jeff0} reads at $R\gg1$ as
\begin{eqnarray}
\label{jeff}
J({\bf r}) 
&=& 
J_0 \frac{ e^{-R} }{R^{3/2}},\\
\label{j0}
J_0 &=& \frac{(4\pi)^{5/2} C}{6\sqrt2} \frac{H_m^2}{V_\xi\Delta} .
\end{eqnarray}
It is seen from Eq.~\eqref{jeff} that $V_\xi$ is a volume of an ellipsoid with axes $\xi_x$, $\xi_y$, and $\xi_z$ inside which the effective exchange coupling is not exponentially small.

We assume for the beginning that spins are classical and consider the role of quantum effects at the end of this section. To estimate the transition temperature $T_N(n)$ in a mean-field manner, we follow Ref.~\cite{Korenblit1973}, start with a very small $n$, and introduce the quantity $R \left( T \right)$ at which 
\begin{equation}
\label{sjt}
S^2J ({\bf r}) = T
\end{equation}
(see  Eq.~\eqref{jeff}). The latter equality determines the surface of an ellipsoid with axes $R \left( T \right)\xi_x$, $R \left( T \right)\xi_y$, $R \left( T \right)\xi_z$ and with the spin at the center. Due to the exponential dependence of $J({\bf r})$ on $\bf r$ and thermal fluctuations, another spin lying inside and outside of the ellipsoid is correlated and uncorrelated with the spin at the center, respectively (provided that other spins are away from these two). Consequently, our task is reduced to the problem of finding a percolation threshold in a system of oriented randomly distributed ellipsoids. The solution of the latter problem is well known: \cite{ShklovskyEfros1979} the infinite network and the percolation disappear when
\begin{equation}
\label{perc}
	nV=2.736,
\end{equation}
where $V = R(T)^3V_\xi$ is the ellipsoids volume. This line of argument is valid as long as $R (T)\agt1$ (because the effective exchange coupling \eqref{jeff0} is not exponentially small at $R<1$) which implies that $V_\xi n \alt 2.7$ (see Eq.~\eqref{perc}). Another restriction appears from the requirement that the ellipsoid with axes $R \left( T \right)\xi_x$, $R \left( T \right)\xi_y$, and $R \left( T \right)\xi_z$ should cover more than one lattice site in each direction so that $R (T)\agt \max\{1,1/\xi_x,1/\xi_y,1/\xi_z\}$. Finding $R(T)$ from Eq.~\eqref{perc} and substituting the result to Eq.~\eqref{jeff}, one obtains from Eq.~\eqref{sjt}
\begin{equation}
\label{tc<}
T_N(n) \propto J_0S^2 \sqrt{V_\xi n} e^{-1.4/(V_\xi n)^{1/3}},
\quad
\mbox{ if } 
\quad
(V_\xi n)^{1/3} \alt \min\{1,\xi_x,\xi_y,\xi_z\},
\end{equation}
where we omit a numerical factor bearing in mind that this is the mean-field estimation of the critical temperature. 

At larger concentration $n$, many scenarios arise depending on values of $\xi_x$, $\xi_y$, and $\xi_z$. We consider now some of them to illustrate the main ideas. At $\xi_x,\xi_y,\xi_z\agt1$ (and at $(V_\xi n)^{1/3} \agt 1$, see Eq.~\eqref{tc<}), ellipsoids with the volume $V_\xi$ form a three-dimensional percolating network and each spin has on average $V_\xi n \agt 1$ neighbors inside the volume $V_\xi$ who interact with it by the exchange coupling of the order of $J_0$ (see Eqs.~\eqref{jeff0} and \eqref{j0}). Then, the mean molecular field acting on the given spin is estimated as $J_0(V_\xi n)$ and it determines the transition temperature in this "three-dimensional" regime that reads as
\begin{equation}
\label{tc>1}
T_N(n) \propto J_0S^2 (V_\xi n),
\quad
\mbox{ if } 
\quad
(V_\xi n)^{1/3} \agt 1
\mbox{ and } 
\xi_x,\xi_y,\xi_z\agt1.
\end{equation}

If some of $\xi_x$, $\xi_y$, and $\xi_z$ is smaller than unity, the ellipsoid with axes $R \left( T \right)\xi_x$, $R \left( T \right)\xi_y$, and $R \left( T \right)\xi_z$ does not cover more than one lattice site in the corresponding direction or directions when $(V_\xi n)^{1/3} \agt \min\{1,\xi_x,\xi_y,\xi_z\}$ (see Eq.~\eqref{tc<}). Let us discuss a "two-dimensional" regime with $\xi_x\ll\xi_y\sim\xi_z<1$. At $(V_\xi n)^{1/3} \agt\xi_x$, the exchange coupling between spins inside the $yz$ plane is much larger than that along the $x$ direction. Then, we have a quasi-2D spin system, the transition temperature of which is determined (up to a logarithmic factor) by the in-plane exchange coupling. The latter is given by Eq.~\eqref{jeff}, where now ${\bf R} = {\bf R}_2 = \left(\frac{y}{\xi_y}, \frac{z}{\xi_z} \right)$. Following the same logic as above, one has to solve a two-dimensional problem of percolation in the system of randomly distributed oriented ellipses with concentration $n$. The percolation arises in this case at \cite{ShklovskyEfros1979} $nV_2=4.51$ (cf.\ Eq.~\eqref{perc}), where $V_2=R_2(T)^2 v_\xi$ and $v_\xi=\pi\xi_y\xi_z$ is the ellipses area. One estimates with the logarithmic precision (cf.\ Eq.~\eqref{tc<})
\begin{equation}
\label{tc>2}
T_N(n) \propto J_0S^2 \xi_x (v_\xi n)^{3/4} e^{-2.1/\sqrt{v_\xi n}},
\quad
\mbox{ if } 
\quad
(V_\xi n)^{1/3} \gg \xi_x
\mbox{ and } 
\xi_x\ll\xi_y\sim\xi_z<1,
\end{equation}
where the factor $\xi_x$ comes from the logarithm of the ratio of the in-plane exchange coupling ($\sim e^{-2.1/\sqrt{v_\xi n}}$) and the inter-plane one ($\sim e^{-1/\xi_x}$).

The most pronounced quantum effect which can influence the results obtained above is the formation of the "non-magnetic" singlet state of two closest spins-$1/2$ coupled by the antiferromagnetic exchange. However we expect that this effect is small in "$d$-dimensional" regimes with $d\ge2$. Indeed, the fraction of spins involved in such couples is estimated as $n^{1-1/d}$ which is much smaller than unity at $d\ge2$.
\footnote{
The probability to find a couple of spins a distance $r$ away from each other who have no closer neighbors is estimated as $n^2(1-n)^{vr^3}\approx n^2e^{-nvr^3}$, where $v$ is a constant of the order of unity giving the volume of two intersecting spheres of a unit radius the distance between centers of which is equal to unity. We use here that $n$ is the probability to find a spin at a given lattice site. Integration of this result on $r$ from unity to infinity gives the total probability to find such couple of spins which is proportional to $n^{2-1/3}$. The generalization of this result to another dimension $d$ reads as $n^{2-1/d}$.
}
Besides, spins interact ferromagnetically in half of these couples.

\section{Low-energy excitations}
\label{excitation}

To lighten notation, we assume below that $\xi_x=\xi_y=\xi_z=\xi$. Corresponding results can be obtained similarly at $\xi_x\ne\xi_y\ne\xi_z$. In particular, general expressions for the specific heat and the magnetization are obtained from those presented below by a simple replacement of $\xi$ by $(\xi_x\xi_y\xi_z)^{1/3}$.

\subsection{Spin waves}
\label{spinwaves}

We discuss first the long-wavelength hydrodynamic excitations (spin waves) at a small concentration of defects $n\ll1$. Such excitations appear due to the disorder-induced magnetically ordered part of the considered system. We will be guided by the linear dispersion relation for low-frequency spin waves \cite{Nakayama},
\begin{equation}
\label{omega}
\omega_{\bf k}=C \left( n \right) k,
\end{equation}  
where the spin-wave velocity $C \left( n \right)$ has the form
\begin{equation}
\label{stiffness1}
C(n) = \sqrt{2\Upsilon(n)/\chi_\perp(n)},
\end{equation}
where $\chi_\perp $ is the transverse susceptibility and $\Upsilon$ is the helicity modulus (i.e., a measure of the energy required to create a spatial variation in the magnetization). The latter quantity can be found as it was done in Ref.~\cite{shenderrev} for disordered ferromagnets. $\Upsilon\propto\sigma/n$, where $\sigma$ is the conductivity of a related resistor network. \cite{Kirkpatrick1973} The electron conductivity is well known \cite{Kurkijarvi1974} in the system of chaotically distributed centers in which the probability of electron jump between centers is determined by Eq.~\eqref{jeff}:
\begin{equation}
\label{sigma}
\sigma\propto r_c^{-(\nu+1)}\frac{1}{r_{c}^{3/2}}e^{-r_c /\xi }, 
\end{equation}
where $\nu$ is the critical index of the correlation length in the percolation theory and 
\begin{equation}
	r_c = 0.87 /n^{1/3}
\end{equation}
 is the critical radius of spheres at which the infinite cluster disappears and which is determined by Eq.~\eqref{perc}. As a result, one obtains from Eq.~\eqref{sigma}
\begin{equation}
\label{helicity}
\Upsilon(n) \propto n^{(2\nu-1)/6} e^{-0.87 /(\xi n^{1/3}) }, 
\end{equation}

It is seen from Eq.~\eqref{helicity} that the main contribution to the helicity modulus is made by those spins whose distance to the nearest neighbors lies in the interval $(r_c-\xi,r_c+\xi)$. This result is natural because the infinite network in which the long-wavelength spin waves can propagate should contain bonds with exchange energy of the order of $J(r_c)$: the coupling energy of spins a distance $r\gg r_c$ away from each other is much smaller than the energy of the spin wave and two spins oscillate in phase if $r\ll r_c$. Because $\xi\ll r_c$, the considered infinite network is the network which arises in a system close to the percolation transition. Then, the correlation length of this network reads as
\begin{equation}
{\cal L} \sim r_c(r_c/\xi)^\nu \propto n^{-(1+\nu)/3}.
\end{equation}
It is well known that the transverse susceptibility of randomly depleted antiferromagnet diverges near the percolation threshold $p_c$ as $(p-p_c)^{-\tau}$. \cite{harkir} Thus, one has for this quantity in our system
\begin{equation}
\label{chi}
\chi_{\perp}(n)
\propto 
(r_{c}/\xi)^{\tau}\left(\frac{\exp\left(-r_c /\xi \right)}{r_{c}^{3/2}}\right)^{-1}
 \propto n^{-(2\tau+3)/6} e^{0.87 /(\xi n^{1/3}) }.
\end{equation}
One obtains from Eqs.~\eqref{stiffness1}, \eqref{helicity}, and \eqref{chi} for the spin-wave velocity
\begin{equation}
\label{stiffness2}
C\left(n\right)\propto n^{(1+\nu+\tau)/6} e^{-0.87 /(\xi n^{1/3})}.
\end{equation}

It is interesting to note that the concentration dependence of $C\left(n\right)/T_N(n)$ does not contain the exponential factor (see Eqs.~\eqref{tc<} and \eqref{stiffness2})
\begin{equation}
C(n)/T_N(n) \propto  n^{(\nu+\tau-2)/6}.
\end{equation}

Propagating spin waves exist in depleted antiferromagnets if their wavelength is larger than the correlation length. \cite{Nakayama} Thus, well-defined spin waves having spectrum \eqref{omega} exist in our system up to the energy
\begin{equation}
\label{wm}
\omega_m \sim C(n)/{\cal L} \propto n^{(3+3\nu+\tau)/6} e^{-0.87 /(\xi n^{1/3})}.
\end{equation}
Excitations with higher energies are localized.

\subsection{Localized excitations}
\label{localexc}

To make further consideration more compact, we omit for simplicity $1/R^{3/2}$ in Eq.~\eqref{jeff} and assume that
\begin{equation}
\label{jeff2}
J({\bf r}) 
=
J_0 e^{-R}
\end{equation}
bearing in mind that the exponential behavior of the effective coupling plays the major role on long distances at $n\ll1$.

As in disordered ferromagnets, \cite{Korenblit,shenderrev} a substantial part of the low-energy spectrum in our system consists of local excitations. The simplest excitation of this type is a local flip of a spin whose nearest neighbor is situated at a distance larger than the average distance $1/n^{1/3}$. Due to the exponential dependence of the exchange coupling, such spins are weakly bound to the bulk of magnetic atoms. The density of states of such excitations is determined by the distribution function $W(\epsilon)$ of molecular fields which can be found in the mean-field approximation as it was done in Refs.~\cite{Korenblit,shenderrev} for ferromagnets. Because the mean-field treatment of antiferromagnets is similar in many respects to that of ferromagnets, we present below main formulas and refer the reader to Refs.~\cite{Korenblit,shenderrev} for extra details. 

The molecular field acting on spin $i$ reads as 
\begin{equation}
\label{molfield}
H_{i} = \langle S\rangle\sum\limits_{j=1}^N J\left({\bf r}_{i}-{\bf r}_{j}\right),
\end{equation}
where $\langle S\rangle$ is the mean spin value and $N$ is the total number of impurities. Then, the molecular-field distribution function has the form \cite{Chandrasekar}
\begin{equation}
\label{distrfunc}
W(\epsilon) = \frac{1}{V^N}\int\delta\left(\epsilon-\langle S\rangle \sum_{j=1}^N J\left({\bf r}_{j}\right)\right)d{\bf r}_{1}...d{\bf r}_N,
\end{equation}
where $V$ is the volume of the system. Taking into account Eq.~\eqref{jeff2} and integrating Eq.~\eqref{distrfunc} by parts, we obtain
\begin{eqnarray}
\label{transdistrf}
W(\epsilon) &=& \frac{1}{2\pi}\int\limits_{-\infty}^{\infty} e^{-ip\epsilon- D(p)} dp,\\
\label{Dfunc}
D(p) &=& iv \langle S\rangle J_{0}p
\int\limits_{0}^{1}
\ln^{3}(x) e^{ip\langle S\rangle J_{0}x} dx,
\end{eqnarray}
where $v=\frac{4\pi}{3}n\xi^{3}\ll 1$.

$W(\epsilon)$ can be found also from the following simplified consideration which is in agreement with Eqs.~\eqref{transdistrf} and \eqref{Dfunc}. At not too small $\epsilon$, when $3v\ln^{2}\left(\langle S\rangle J_{0}/\epsilon\right)\ll 1$ and $\epsilon\ll \langle S\rangle J_{0}$, the distribution function is determined by molecular fields acting on spins whose distances to all other spins are larger than the average distance $1/n^{1/3}$. The molecular field $\epsilon$ acting on such a spin and made by its nearest neighbor located at distance $r(\epsilon)$ reads as (see Eq.~\eqref{jeff2}) 
\begin{equation}
\label{distance}
\epsilon =  \langle S\rangle J_{0} \exp\left(-r(\epsilon)/\xi\right)
\end{equation}
Each spin located in a spherical layer (with the considered spin at the center) of radius $r(\epsilon)$ and thickness $\xi$ produces this molecular field. Because the number of unpaired spins given by $4\pi nr^{2}(\epsilon)\xi$ is small in this layer in the considered regime ($3v\ln^{2}\left(\langle S\rangle J_{0}/\epsilon\right)\ll 1$), the molecular field distribution function is determined by the probability to find a nearest neighbor at distance $r(\epsilon)$ which is given by the Poisson distribution \cite{Chandrasekar}
\begin{equation}
\label{poisson}
W(\epsilon) = W_P(\epsilon) 
= 
4\pi nr^{2}(\epsilon)\exp\left(-\frac{4\pi}{3}nr^{3}(\epsilon)\right)
\left| \frac{dr}{d\epsilon} \right|
= 
\frac{3v}{\epsilon} 
\ln^{2} \left( \frac{\langle S\rangle J_{0}}{\epsilon} \right)
\exp \left( -v\ln^{3} \left( \frac{\langle S\rangle J_{0}}{\epsilon}  \right) \right).
\end{equation}
$W(\epsilon)$ given by Eq.~\eqref{poisson} grows as $\epsilon$ decreases approximately as $1/\epsilon$ when $v\ln^{3}\left(\langle S\rangle J_{0}/\epsilon\right)\ll 1$.

At smaller molecular fields, when $3v\ln^{2}\left(\langle S\rangle J_{0}/\epsilon\right)\gg 1$, the number is large of spins in the layer with radius $r(\epsilon)$ and thickness $\xi$. Then, $W(\epsilon)<W_P(\epsilon)$ in this regime and, consequently, $W(\epsilon)\to0$ at $\epsilon\to0$ (see Eq.~\eqref{poisson}). Thus, $W(\epsilon)$ has a maximum at $\epsilon_m$ satisfying $3v\ln^{2}\left(\langle S\rangle J_{0}/\epsilon_m\right)\approx 1$. 

Because $W(\epsilon)$ tends to zero as $\epsilon\to 0$ faster than any power law, spin waves give the major contribution to the density of states at $\epsilon<\omega_m$, where $\omega_m$ is given by Eq.~\eqref{wm}.  

\section{Magnetization and specific heat}
\label{magnspecheat}

The specific heat can be expressed as follows: \cite{Korenblit}
\begin{equation}
\label{specheat1}
C_{m}=n\int\limits_{0}^{\infty}C\left(\epsilon\right)W\left(\epsilon\right)\,d\epsilon,
\end{equation}
\begin{equation}
\label{specheat2}
C\left(\epsilon\right)
=
\left(\beta\epsilon\right)^{2}\left[\frac{1}{4\sinh^{2}(\beta\epsilon/2)}
-
\frac{\left(S +1/2\right)^{2}}{\sinh^{2}\beta\epsilon(S +1/2)}\right],
\end{equation} 
where $\beta=1/T$. Eq.~\eqref{specheat1} may be rewritten as
\begin{eqnarray}
\label{specheat3}
C_{m}
&=&
n\int\limits_{0}^{\infty} 
W\left(\epsilon\right)
F(\epsilon)
\left( \beta\epsilon \right)^{2}
e^{-\beta\epsilon}
\,d\epsilon,\\
\label{F}
F(\epsilon)
&=&
\frac{e^{2\beta\epsilon}}{\left(e^{\beta\epsilon}-1\right)^{2}}-\frac{\left(2S+1\right)^{2}e^{2\left(S+1\right)\beta\epsilon}}{\left(e^{\left(2S+1\right)\beta\epsilon}-1\right)^{2}}.
\end{eqnarray} 
The bounded smooth function $F(\epsilon)$ is positive at $\epsilon\ge0$, $F(0)=\frac{S(S+1)}{3}$, and $F(\infty)=1$. To estimate the integral in Eq.~\eqref{specheat3}, let us consider first the following quantity (cf.\ Eq.~\eqref{specheat3}):
\begin{equation}
\label{I1}
I_{1}=\beta^{2}\int\limits_{0}^{\infty}W\left(\epsilon\right)\epsilon^{2}e^{-\beta\epsilon}\,d\epsilon=\beta^{2}\frac{d^2}{d\beta^2}\int\limits_{0}^{\infty}W\left(\epsilon\right)e^{-\beta\epsilon}\,d\epsilon.
\end{equation}
One obtains from Eqs.~\eqref{transdistrf} and \eqref{Dfunc}
\begin{eqnarray}
\label{I0}
I_{0}
&=&
\int\limits_{0}^{\infty}W\left(\epsilon\right)e^{-\beta\epsilon}\,d\epsilon
=
e^{-D\left(i\beta\right)}
=
e^{-vf\left(\beta SJ_{0}\right)},\\
\label{f}
f(x)
&=&
(\gamma+\ln x)^{3}+\frac{\pi^2}{2}(\gamma+\ln x) + 2\zeta(3),
\end{eqnarray}
where $\gamma$ is the Euler constant and $\zeta(x)$ is the zeta-function. Substituting Eq.~\eqref{I0} into Eq.~\eqref{I1}, one obtains
\begin{equation}
\label{I1new}
I_{1}
=
\left[(vf_{1})^{2}+vf_{1}-6v(\ln(\beta SJ_{0})+\gamma)\right] e^{-vf(\beta SJ_{0})},
\end{equation}
where $f_{1}=3(\ln(\beta SJ_{0})+\gamma)^{2}+\pi^{2}/2$. It is clear from Eqs.~\eqref{I1}, \eqref{I0}, and \eqref{I1new} that $I_1\gg I_0$ at small temperature when $3v\ln^{2}\left(\beta SJ_{0}\right)\gtrsim 1$ (i.e., when $T\lesssim SJ_{0}\exp(-1/\sqrt{3v})$). Hence, the main contributions to the integral in Eq.~\eqref{specheat3} comes from $\epsilon\beta\gg 1$ in which case one can replace $F(\epsilon)$ by $F(\infty)=1$. Then, one obtains for the specific heat
\begin{equation}
\label{specheat4}
C_{m}=n\left[(vf_{1})^{2}+vf_{1}-6v(\ln(\beta SJ_{0})+\gamma)\right] e^{-vf(\beta SJ_{0})}
\quad
\mbox{at}
\quad
3v\ln^{2}\left(\beta SJ_{0}\right)\gtrsim 1.
\end{equation}   
In the opposite limiting case of $3v\ln^{2}\left(\beta SJ_{0}\right)\lesssim 1$, the consideration becomes somewhat more involved. It can be carried out using series expansion of $F(\epsilon)$ in powers of $e^{-\beta \epsilon}$ as it is done in Ref.~\cite{Korenblit}, the result being
\begin{equation}
\label{specheat5}
C_{m} = nvS(S+1)\ln^{2}(\beta SJ_{0})\exp(-v\ln^{3}(\beta SJ_{0}))
\quad
\mbox{at}
\quad
3v\ln^{2}\left(\beta SJ_{0}\right)\lesssim 1.
\end{equation} 

The average impurity spin is given by \cite{Korenblit}
\begin{equation}
\label{magnetization}
\langle S^{z}\rangle =\int\limits_{0}^{\infty}W(\epsilon)\left[(S+1/2)\coth((S+1/2)\epsilon\beta)-\frac{1}{2}\coth(\epsilon\beta/2)\right]\,d\epsilon.
\end{equation}
Representing $\coth((S+1/2)\epsilon\beta)$ and $\coth(\frac{\epsilon\beta}{2})$ as series in powers of $e^{-\beta \epsilon}$, the calculation is reduced to taking integrals of the type \eqref{I0} with the result \cite{Korenblit}
\begin{equation}
\label{magn1}
\langle S^{z}\rangle = S-\exp[-vf(\beta SJ_{0})]
\quad
\mbox{at}
\quad
v\ln^3\left(\beta SJ_{0}\right)\gtrsim 1
\end{equation}
and
\begin{equation}
\label{magn2}
\langle S^{z}\rangle = S[1-\exp(-v\ln^{3}(\beta SJ_{0}))]
\end{equation}
at $3v\ln^{2}\left(\beta SJ_{0}\right)\lesssim 1$ and $v\ln^3\left(\beta SJ_{0}\right)\gtrsim 1$. Notice that the inequality $v\ln^{3}\left(\beta SJ_{0}\right)\gtrsim 1$ corresponds to the condition $T\lesssim T_{N}$ which assumes $S-\langle S^{z}\rangle\ll S$. At small temperatures when $3v\ln^{2}\left(\beta SJ_{0}\right)\gtrsim 1$, magnetization \eqref{magn1} drops off faster than any power law. It means that $\langle S^{z}\rangle$ is basically governed by spin waves at such $T$ in agreement with conclusions of Sec.~\ref{localexc}.

Owing to the pre-exponential factor in \eqref{specheat4} and \eqref{specheat5}, the specific heat falls off with decreasing temperature more slowly than the magnetization does. As a result, the role of local unpaired-spin flips is more essential in the specific heat as opposed to the deviation of the average spin from the saturation value. 

\section{Application to relevant compounds}
\label{experiment}

\begin{figure}
\includegraphics[scale=0.4]{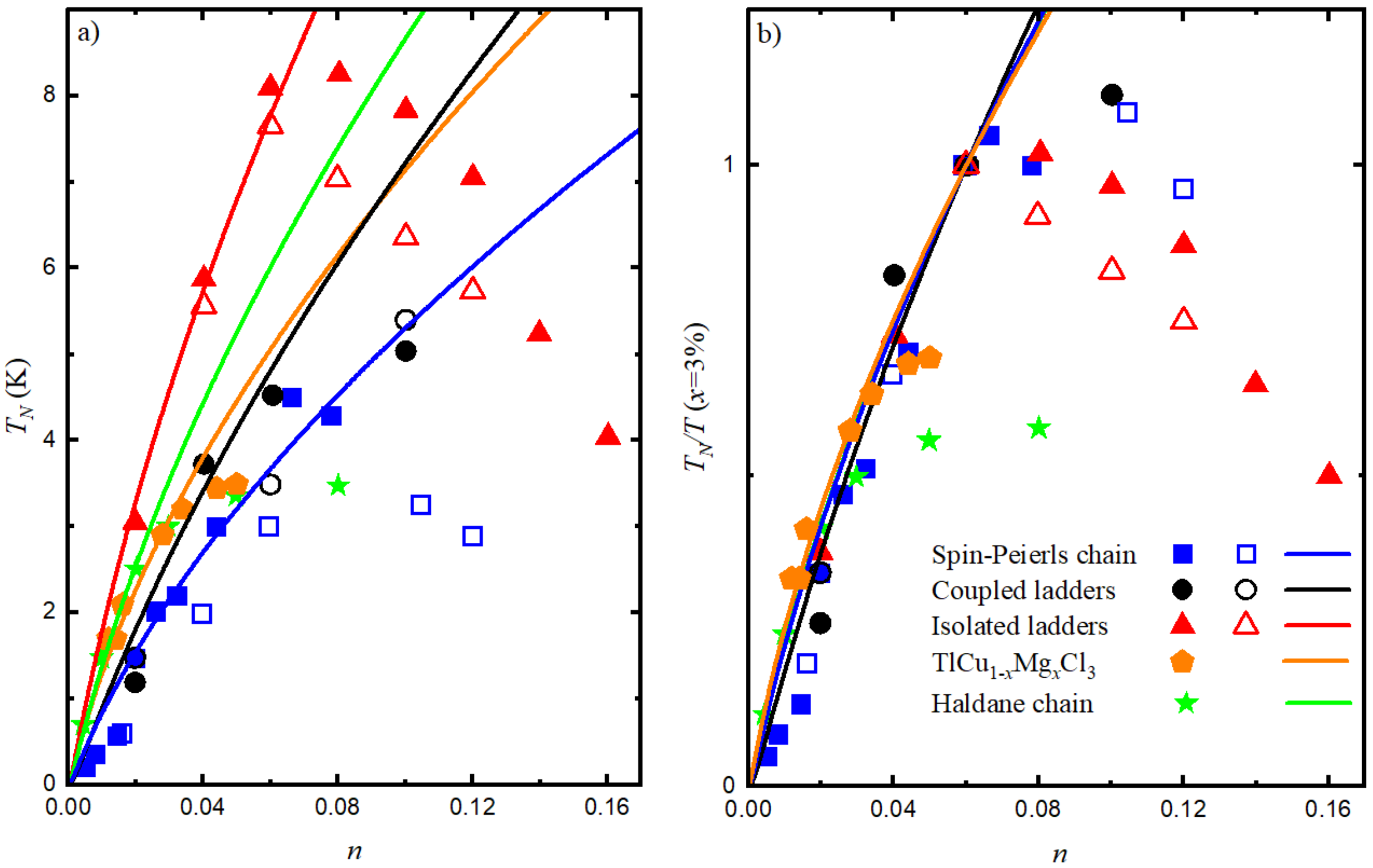}
\caption{a) N\'eel temperature $T_N$ as a function of concentration of unpaired spins $n$ in doped spin-gapped systems $\rm TlCu_{1-{\it x}}Mg_{{\it x}}Cl_{3}$ (Refs.~\cite{Oosawa2002, Suzuki2011}), spin ladders $\rm Bi(Cu_{1-{\it x}}(Zn\mbox{ or }Ni)_{{\it x}})_{2}PO_{6}$ (Ref.~\cite{bobroff}) and $\rm Sr(Cu_{1-{\it x}}(Zn\mbox{ or }Ni)_{{\it x}})_{2}O_{3}$ (Refs.~\cite{ladder3,f3}), Haldane chain materials $\rm Pb(Ni_{1-{\it x}}Mg_{{\it x}})_{2}V_{2}O_{8}$ (Refs.~\cite{haldane4, haldane5}), and spin Peierls chain compounds $\rm Cu_{1-{\it x}}(Zn\mbox{ or }Ni)_{{\it x}}GeO_{3}$ (Ref.~\cite{peierls2}). The full and open symbols correspond to nonmagnetic and magnetic impurities, respectively. Presented experimental data were taken from the cited papers by digitizing corresponding plots in them. Solid lines are drawn using Eq.~\eqref{tc<} with parameters summarized in Table~\ref{tab:ksi} (the constant of proportionality in Eq.~\eqref{tc<} is fitted for each material). b) Same data as in a) but $T_N$ is divided by its value at $x=3\%$.
\label{fig1}}
\end{figure}

The theory developed above can be applied to the following gapped compounds doped with magnetic and non-magnetic impurities a lot of experimental data for which have been obtained so far: spin-$\frac12$ dimer system $\rm TlCu_{1-{\it x}}Mg_{{\it x}}Cl_{3}$, spin-ladder materials $\rm Bi(Cu_{1-{\it x}}Zn_{{\it x}})_{2}PO_{6}$ and $\rm Sr(Cu_{1-{\it x}}Zn_{{\it x}})_{2}O_{3}$, spin-Peierls chain $\rm Cu_{1-{\it x}}Zn_{{\it x}}GeO_{3}$, and spin-1 Haldane chain $\rm Pb(Ni_{1-{\it x}}Mg_{{\it x}})_{2}V_{2}O_{8}$. Parameters of these substances are collected in Table~\ref{tab:ksi}. As it is seen from Fig.~\ref{fig1}(a), the transition temperature $T_N$ is described well by Eq.~\eqref{tc<} in all of these spin systems at $n<0.06$. Fig.~\ref{fig1}(b) demonstrates that $T_N(n)/T_N(x=3\%)$ given by Eq.~\eqref{tc<} shows a linear-like behavior in the considered range of $n$ which describes well the experimental data. The seeming universality of $T_N(n)/T_N(x=3\%)$ in the considered compounds at $n<0.06$ was noted first in Ref.~\cite{bobroff}. It is seen also from Fig.~\ref{fig1} that a deviation of theoretical curves from experimental points takes place outside of the domain of Eq.~\eqref{tc<} validity (i.e., at $(V_\xi n)^{1/3}\agt1$), where a one-dimensional behavior is expected from the above discussion. Consideration of this regime is out of the scope of the present paper.

\begin{figure}
\includegraphics[scale=0.35]{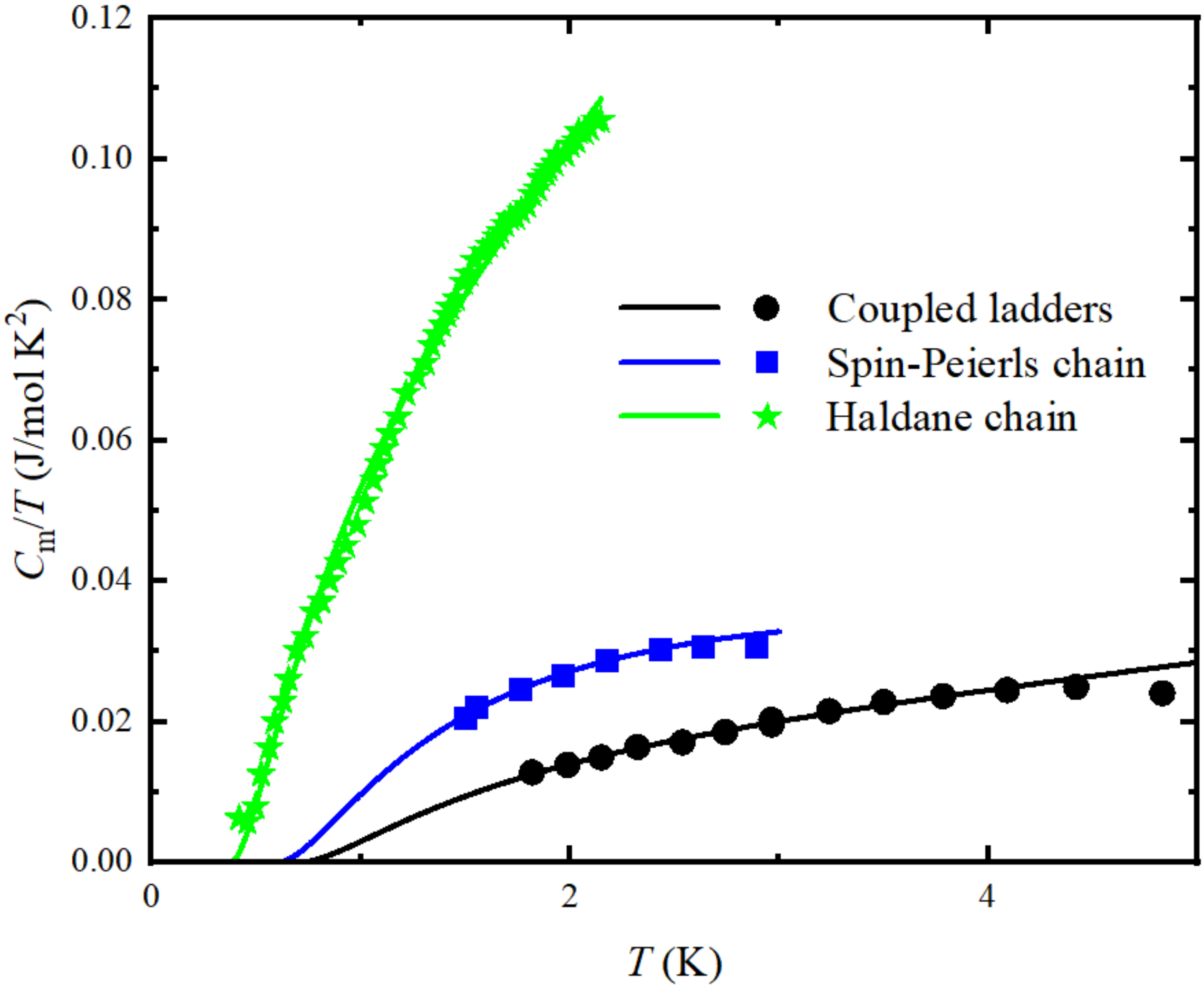}
\caption{Magnetic specific heat $C_{m}$ divided by $T$ for doped two-leg spin ladder $\rm Bi(Cu_{0.97}Zn_{0.03})_{2}PO_{6}$ (Ref.~\cite{Koteswararao2010}), Haldane chain materials $\rm Pb(Ni_{0.98}Mg_{0.02})_{2}V_{2}O_{8}$ (Refs.~\cite{Masuda2002}), and spin-Peierls chain compounds $\rm Cu_{0.98}Zn_{0.02}GeO_{3}$ (Ref.~\cite{Oseroff1995}). Solid lines are drawn using Eq.~\eqref{specheat5} (with the replacement of $\xi$ by $(\xi_x\xi_y\xi_z)^{1/3}$ and with parameters from Table~\ref{tab:ksi}) and varying $J_0$ and the overall constant.
\label{fig2}}
\end{figure}

We are not aware of experimental results for the magnetization. The magnetic part of the specific heat was measured before in $\rm Bi(Cu_{0.97}Zn_{0.03})_{2}PO_{6}$ (Ref.~\cite{Koteswararao2010}), $\rm Cu_{0.98}Zn_{0.02}GeO_{3}$ (Ref.~\cite{Oseroff1995}), and $\rm Pb(Ni_{0.98}Mg_{0.02})_{2}V_{2}O_{8}$ (Refs.~\cite{Masuda2002}) at quite large temperatures when Eq.~\eqref{specheat5} is valid. We present the available experimental data in Fig.~\ref{fig2} and demonstrate that they can be fitted accurately by Eq.~\eqref{specheat5} (with the replacement of $\xi$ by $(\xi_x\xi_y\xi_z)^{1/3}$ and with parameters from Table~\ref{tab:ksi}) varying $J_0$ and the overall constant.

\begin{table}
\caption{\label{tab:ksi}
Correlation lengths $\xi_{x,y,z}$ and gap $\Delta$ values in considered spin-gapped compounds found in previous experimental and numerical works.}
\begin{center}
\begin{tabular}{|c|c|c|c|c|c|c|}
\hline
& & $\Delta$(K) & $\xi_{x}$ & $\xi_{y}$ & $\xi_{z}$ & Remarks \\
\hline
 3D dimer compound & $\rm TlCu_{1-{\it x}}Mg_{{\it x}}Cl_{3}$ & 7.5 & 9.6 & 2.5 & 2.5 & Refs.~\cite{tlcucl,tlcucl2,tlcucl3,tlcucl4}, $n=2x$ \\
\hline
 spin ladder & $\rm Bi(Cu_{1-{\it x}}(Zn\mbox{ or }Ni)_{{\it x}})_{2}PO_{6}$ & 35 & 3.9 & 2.1 & 1.5 & Ref.~\cite{bobroff}, $n=2x$ \\
\hline
 spin ladder & $\rm Sr(Cu_{1-{\it x}}(Zn\mbox{ or }Ni)_{{\it x}})_{2}O_{3}$ & 420 & 8.1 & $\sim2$ & $\sim2$ & Refs.~\cite{ladder4,ladder5,bobroff}, $n=2x$ \\
\hline
 spin-1 (Haldain) chain & $\rm Pb(Ni_{1-{\it x}}Mg_{{\it x}})_{2}V_{2}O_{8}$ & 30 & $\sim8$ & $\sim2$ & $\sim2$ & Refs.~\cite{bobroff,haldane4, haldane5}, $n=x$ \\
\hline
 spin-Peierls chain & $\rm Cu_{1-{\it x}}(Zn\mbox{ or }Ni)_{{\it x}}GeO_{3}$ & 23 & $\sim10$ & $\sim3$ & $\sim1$ & Refs.~\cite{peierls3,peierls4,peierls5}, $n=2x$ \\
\hline
\end{tabular}
\end{center}
\end{table} 

\section{Summary and conclusion}
\label{conc}

In conclusion, we discuss theoretically the magnetically ordered phase induced by small concentration $n$ of magnetic and nonmagnetic impurities in gapped three-dimensional and quasi-low-dimensional systems with singlet ground states. We apply the percolation theory to find analytical expressions for the transition temperature $T_N(n)$ to the N\'eel phase, density of low-energy excited states, magnetization and specific heat. The low-energy part of the impurity-induced band of excitations (i.e., the energy interval from zero to $\omega_m$ given by Eq.~\eqref{wm}) is composed of propagating antiferromagnetic spin waves whose spectrum is given by Eqs.~\eqref{omega} and \eqref{stiffness2}. Above spin waves, localized excitations arise. Our expression \eqref{tc<} for $T_N(n)$ describes well available experimental data at $n<0.06$ obtained in spin-$\frac12$ dimer materials, spin-ladder compounds, spin-Peierls and Haldane chain materials (see Fig.~\ref{fig1}). The obtained analytical result \eqref{specheat5} for the magnetic specific heat $C_{m}$ is in good agreement with available experimental findings at $n<0.06$ and $T\leqslant T_{N}$ (see Fig.~\ref{fig2}). 

\begin{acknowledgments}

This work is supported by the Foundation for the Advancement of Theoretical Physics and Mathematics "BASIS" and by RFBR according to the research Project No.\
18-02-00706.

\end{acknowledgments}

\bibliography{bib}

\end{document}